\documentclass[preprint]{imsart}

\usepackage{amsthm,amsmath,amsfonts}
\RequirePackage[colorlinks,citecolor=blue,urlcolor=blue]{hyperref}
\usepackage[authoryear]{natbib}
\usepackage[english]{babel}
\usepackage{latexsym}
\usepackage{subfigure}
\usepackage{epsfig}
\usepackage{color,soul}
\usepackage{moreverb}
\usepackage{mathtools}
\usepackage{tikz, color}
\usepackage{enumerate,linegoal}
\usepackage{calc}
\usepackage{latexsym}
\usepackage{amsmath}
\usepackage{amsfonts}
\usepackage{amssymb}
\usepackage{bm}
\usepackage{psfrag}
\usepackage{graphicx}
\usepackage{epstopdf}
\usepackage{framed}
\usepackage{booktabs}
 \usepackage[flushleft]{threeparttable}

\usepackage{xfrac,array,booktabs}
\usepackage{verbatim}
\usepackage{enumitem}
\usepackage{pdflscape}
\usepackage{rotating}

\usetikzlibrary{matrix}

\startlocaldefs

\renewcommand{\eqref}[1]{(\ref{#1})}

\newcommand{\tp}{^{\top}}

\endlocaldefs

\begin{document}

\begin{frontmatter}

\title{Letter to the Editor}
\runtitle{Letter to the Editor}

\begin{aug}
\author{\fnms{Marco} \snm{Geraci}\corref{}\thanksref{t1}\ead[label=e1]{geraci@mailbox.sc.edu}}

\thankstext{t1}{Department of Epidemiology and Biostatistics, Arnold School of Public Health, University of South Carolina, 915 Greene Street, Columbia SC 29209, USA. \printead{e1}}

\runauthor{M. Geraci}

\affiliation{University of South Carolina\thanksmark{t1}}

\end{aug}

\begin{abstract}
\quad \cite{galarza} have recently proposed a method of estimating linear quantile mixed models \citep{geraci_2014b} based on a Monte Carlo EM algorithm. They assert that their procedure represents an improvement over the numerical quadrature and non-smooth optimization approach implemented by \cite{geraci_2014a}. The objective of this note is to demonstrate that this claim is incorrect. We also point out several inaccuracies and shortcomings in their paper which affect other results and conclusions that can be drawn.
\end{abstract}



\end{frontmatter}

\section{Linear quantile mixed models}
Linear quantile mixed models (LQMMs) were developed by \cite{geraci_2014b} as an extension of the quantile regression model with random intercepts of \cite{geraci_2007}. We consider data from two-level nested designs in the form $(\mathbf{x}_{ij}\tp,\mathbf{z}_{ij}\tp,y_{ij})$, for $j=1,\ldots , n_{i}$ and $i=1,\ldots , M$, $N = \sum_i n_i$, where $\mathbf{x}_{ij}\tp$ is the $j$th row of a known $n_{i}\times p$ matrix $\mathbf{X}_i$, $\mathbf{z}_{ij}\tp$ is the $j$th row of a known $n_{i}\times q$ matrix $\mathbf{Z}_i$ and $y_{ij}$ is the $j$th observation of the response vector $\mathbf{y}_i = (y_{i1},\ldots,y_{in_{i}})\tp$ for the $i$th cluster. The $N\times 1$ vector of responses is denoted by $\mathbf{y} = (\mathbf{y}_{1}\tp,\ldots,\mathbf{y}_{M}\tp)\tp$. This kind of data arise from longitudinal or panel studies and other cluster sampling designs.

The $\tau$th LQMM is defined as
\begin{equation}\label{eq:1}
Q_{y_{ij}|\mathbf{u}_{i}}(\tau) = \mathbf{x}_{ij}\tp\bm\beta_{\tau} + \mathbf{z}_{ij}\tp\mathbf{u}_{i},
\end{equation}
where $0 < \tau < 1$ is the given quantile level, $\bm\beta_{\tau}$ is a $p \times 1$ vector of $\tau$-specific coefficients that are common to all clusters, while the $q \times 1$ vector $\mathbf{u}_{i}$ may vary with cluster. For estimation purposes only, \cite{geraci_2014b} introduced the convenient assumption that the responses $y_{ij}$, $j=1,\ldots , n_{i}$, $i=1,\ldots,M$, conditionally on a $q\times1$ vector of random effects $\mathbf{u}_i$, independently follow the asymmetric Laplace (AL) density
\begin{equation}\label{eq:2}
p(y_{ij}|\mathbf{u}_{i}) = \frac{\tau(1-\tau)}{\sigma_{\tau}}\exp\left\{-\frac{1}{\sigma_{\tau}}\rho_\tau\left(y_{ij}-\mu_{\tau,ij}\right)\right\},
\end{equation}
where $\rho_\tau(r)=r\left\{\tau-I(r < 0)\right\}$ is the `check' function and $I$ denotes the indicator function, with location and scale parameters given by $\mu_{\tau,ij} = \mathbf{x}_{ij}\tp\bm\beta_{\tau} + \mathbf{z}_{ij}\tp\mathbf{u}_{i}$ and $\sigma_{\tau}$, respectively, which we write as $y_{ij} \sim \mathcal{AL}\left(\mu_{\tau,ij}, \sigma_{\tau}\right)$. (The third parameter of the AL is the skew parameter $\tau \in (0,1)$ which, in this model, is fixed and defines the quantile level of interest.) Also, they assumed that $\mathbf{u}_{i} = \left(u_{i1},\ldots,u_{iq}\right)\tp$, for $i=1,\ldots,M$, is a random vector independent from the model's error term with mean zero and $\tau$-specific variance-covariance matrix $\bm\Sigma_{\tau}$ of dimensions $q\times q$. The latter is reparameterized in terms of an $m$-dimensional vector, $1 \leq m \leq q(q+1)/2$, of non-redundant parameters $\bm\theta_{\tau}$, i.e. $\bm\Sigma_{\tau} = \bm\Sigma(\bm\theta_{\tau})$.

The algorithm to estimate $(\bm\beta_{\tau}, \bm\theta_{\tau}, \sigma_{\tau})$ is described in detail by \cite{geraci_2014a,geraci_2014b}. First, the (quasi) log-likelihood is integrated numerically over the distribution of the random effects, i.e.
\begin{align}\label{eq:3}
& \ell_{\mathrm{GQ}}(\bm\beta_{\tau}, \bm\theta_{\tau}, \sigma_{\tau}|\mathbf{y}) = \\
\nonumber & \quad \sum_{i}^{M}\log\left\{\sum_{k_1=1}^{K}\cdots\sum_{k_q=1}^{K}p\left(\mathbf{y}_{i}| \mathbf{v}_{k_1,\ldots,k_q}\right) \prod_{l=1}^{q}w_{k_{l}}\right\},
\end{align}
with $\mathbf{v}_{k_1,\ldots,k_q}=(v_{k_1},\ldots,v_{k_q})\tp$, where $v_{k_l}$ and $w_{k_l}$, $k_l = 1,\ldots,K$, $l=1,\ldots,q$, denote the abscissas and the weights of the (one-dimensional) Gaussian quadrature. Second, the integrated log-likelihood \eqref{eq:3} is maximized via a non-smooth optimizer. In principle, one can consider different distributions for the random effects, which may be naturally linked to different quadrature rules (or penalties). For example, it is immediate to verify that the normal distribution is akin to a Gauss-Hermite quadrature, a special case of LQMM discussed by Geraci and Bottai.

In their paper, Galarza et al write
\vskip 0.2cm
\begin{small}
[Geraci and Bottai (2014)] extended [Geraci and Bottai's (2007)] setup to accommodate multiple random effects [...]. Here, we consider a more general correlated random effects framework with general dispersion matrix $\bm\Psi = \bm\Psi(\mathbf{a})$.
\end{small}
\vskip 0.2cm
This statement is not justified since their model is precisely the LQMM defined in \eqref{eq:1}-\eqref{eq:2} with normal random effects and general variance-covariance matrix $\bm\Sigma_{\tau} = \bm\Sigma(\bm\theta_{\tau})$.

\section{Simulation study}
To fit the LQMM \eqref{eq:1}-\eqref{eq:2}, Galarza et al proposed to use a stochastic approximation of the expectation-maximization (SAEM) algorithm. They compared their estimation approach, as implemented in the \texttt{qrLMM} package, to the quadrature-based algorithm implemented in the \texttt{lqmm} package \citep{geraci_2014a}. The description of their simulation study is not clear as it lacks several details. First of all, it does not specify which versions of \texttt{lqmm}, \texttt{qrLMM} or \texttt{R} were used in their study. Most importantly, there is no indication about \texttt{lqmm} optimization settings and the syntax used for modeling. We speculate that the default options were used. We tried to replicate their design and we believe that the setting described below is quite similar to theirs.

We generated data according to the model
\begin{equation}\label{eq:4}
y_{ij} = \mathbf{x}_{ij}\tp\bm\delta + \mathbf{z}_{ij}\tp\mathbf{u}_{i} + \varepsilon_{\tau,ij},
\end{equation}
where $\bm\delta = (0.8,0.5,1)\tp$, $x_{1,ij} = 1$, $x_{2,ij} \sim \mathcal{N}(0,1)$, $x_{3,ij} \sim \mathcal{N}(0,1)$, $z_{1,ij} \sim \mathcal{N}(0,1)$, $z_{2,ij} \sim \mathcal{N}(0,1)$, and $\varepsilon_{\tau,ij}\sim \mathcal{AL}\left(0, 0.2\right)$. Moreover, $\mathbf{u}_{i} \sim \mathcal{N}(\mathbf{0},\bm\Sigma)$, where
\[
\bm\Sigma = \left[
           \begin{array}{cc}
             0.8 & 0.5 \\
             0.5 & 1 \\
           \end{array}
         \right].
\]
The number of clusters $M$ varied (50, 100, 200, 300) while the size of the clusters was fixed to $n_{i} = 3$, $i = 1,\ldots,M$, throughout the simulation. We considered five LQMMs \eqref{eq:1} for $\tau \in \{0.05, 0.1, 0.5, 0.9, 0.95\}$. (Note that the error $\varepsilon_{\tau,ij}$ in \eqref{eq:4} is sampled from an AL with skewness determined by the same $\tau$ that defines the quantile to be estimated, therefore $\bm\beta_{\tau} = \bm\delta$ for all $\tau$.) Data were replicated 100 times for each combination of sample size and quantile level.

For this simulation, we used \texttt{lqmm} 1.5.3, which is, at the time of writing, the latest version available on the Comprehensive R Archive Network, and \texttt{qrLMM} 1.3 for \texttt{R} version 3.4.2 \citep{R}. By default, the function \texttt{QRLMM} starts the SAEM algorithm with estimates of $\bm\beta_{\tau}$ and $\sigma_{\tau}$ obtained from linear programming (package \texttt{quantreg}). In contrast, \texttt{lqmm} starts by default from ordinary least squares estimates. Therefore, we changed the option \texttt{lqmmControl(startQR = TRUE)} for the sake of comparability. Moreover, we used $K = 9$ quadrature knots instead of the default $K=7$ to improve accuracy since $q > 1$ \citep[see][for details]{geraci_2014b}. For the SAEM algorithm we used the same settings as in Galarza et al, namely 20 Monte Carlo simulations, 500 maximum iterations and $0.2$ for the cut-point that determines the proportion of initial iterations with no memory. The variance-covariance matrix was specified as a general positive-definite matrix in both estimation procedures. All the other estimation settings in \texttt{lqmm} and \texttt{QRLMM} were left unchanged to their default values. In a preliminary analysis, we assessed the computational time needed to run the full simulation and we estimated it would take approximately two months for \texttt{QRLMM}, but less than half an hour for \texttt{lqmm}. Given the excessive computational time needed for \texttt{QRLMM}, we ran the latter only for selected scenarios, namely $M \in\{50, 300\}$ and $\tau \in\{0.05, 0.5, 0.95\}$.

\begin{landscape}
\begin{table}
\caption{Performance summary for the quadrature \& non-smooth optimization algorithm (lqmm) and the approximated EM algorithm (qrLMM) run on a 64-bit operating system machine with 32 Gb of RAM and 3.60 GHz clock-rate processor. All figures refer to the same subset of scenarios, namely $M \in\{50, 300\}$ and $\tau \in\{0.05, 0.5, 0.95\}$. Averages are calculated over $2\times 3 \times 100 = 600$ replicated datasets.}
\label{tab1}
\begin{tabular}{lrrrrr}
\hline
\textit{Algorithm} & \multicolumn{1}{l}{\textit{Average bias}} & \multicolumn{1}{l}{\textit{Average root mean}} & \multicolumn{1}{l}{\textit{Total elapsed time}} & \multicolumn{1}{l}{\textit{Average elapsed time}} & \textit{Percentage of convergence}\\
&\multicolumn{1}{l}{}  & \multicolumn{1}{l}{\textit{squared error}} & \multicolumn{1}{l}{} & \multicolumn{1}{l}{} & \multicolumn{1}{l}{\textit{failures}}\\
\hline
lqmm  &    0.0019 & 0.0852 &    7.5 (min) &  0.7 (s) &  $0\%$ \\
qrLMM &    0.0017 & 0.0910 &   24066.0 (min) &  2551.2 (s) & $21\%$ \\
\hline
\end{tabular}
\end{table}
\end{landscape}

\begin{landscape}
\begin{figure}
\includegraphics[scale = 0.6]{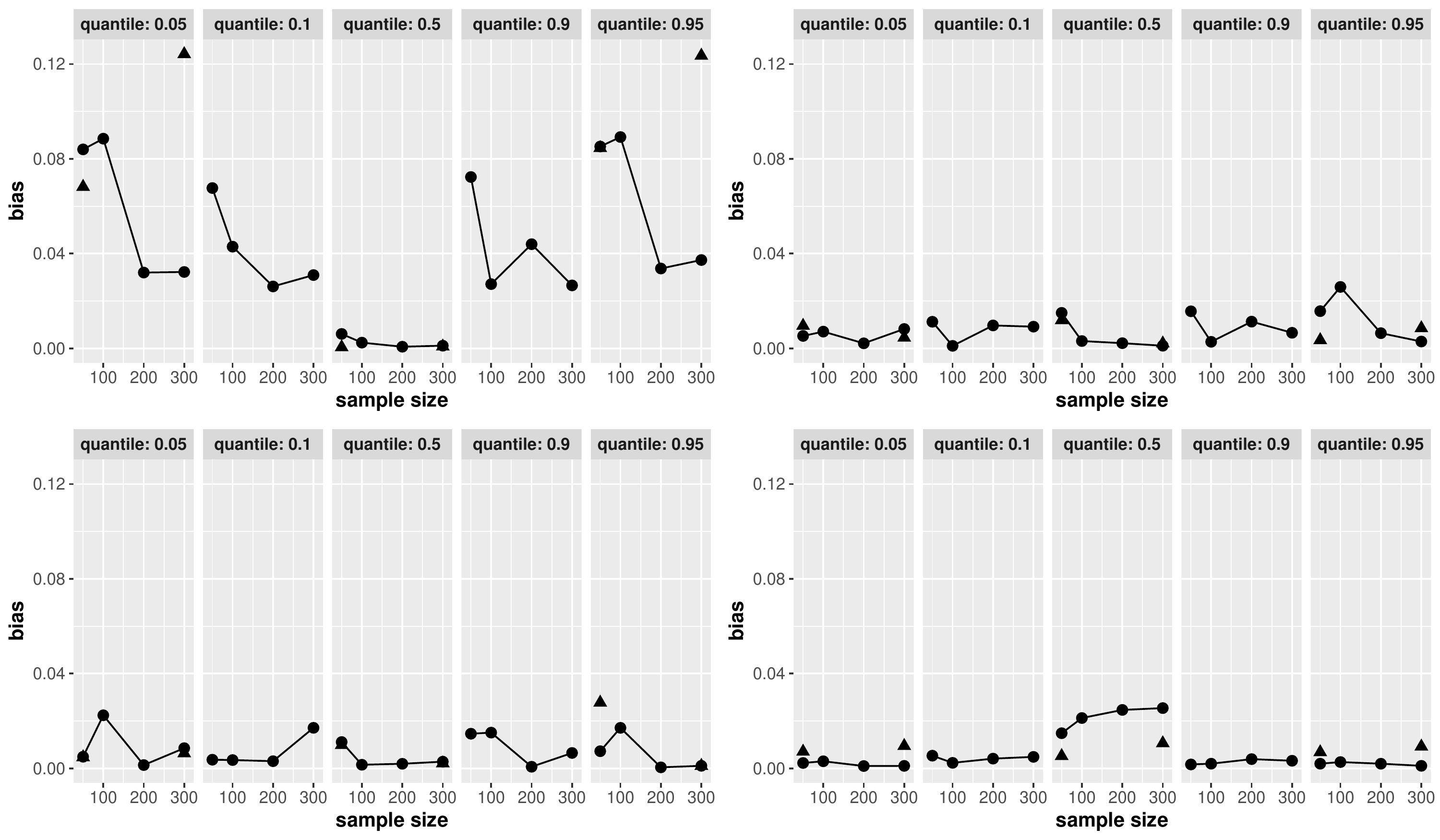}
\caption[]{Absolute values of bias for linear quantile mixed effects estimators based on quadrature \& non-smooth optimization (filled circles) and approximated EM (filled triangles): $\beta_{\tau,0}$ (upper left panels), $\beta_{\tau,1}$ (upper right panels), $\beta_{\tau,2}$ (lower left panels), and $\sigma_{\tau}$ (lower right panels).}
\label{fig1}
\end{figure}
\end{landscape}

\begin{landscape}
\begin{figure}
\includegraphics[scale = 0.6]{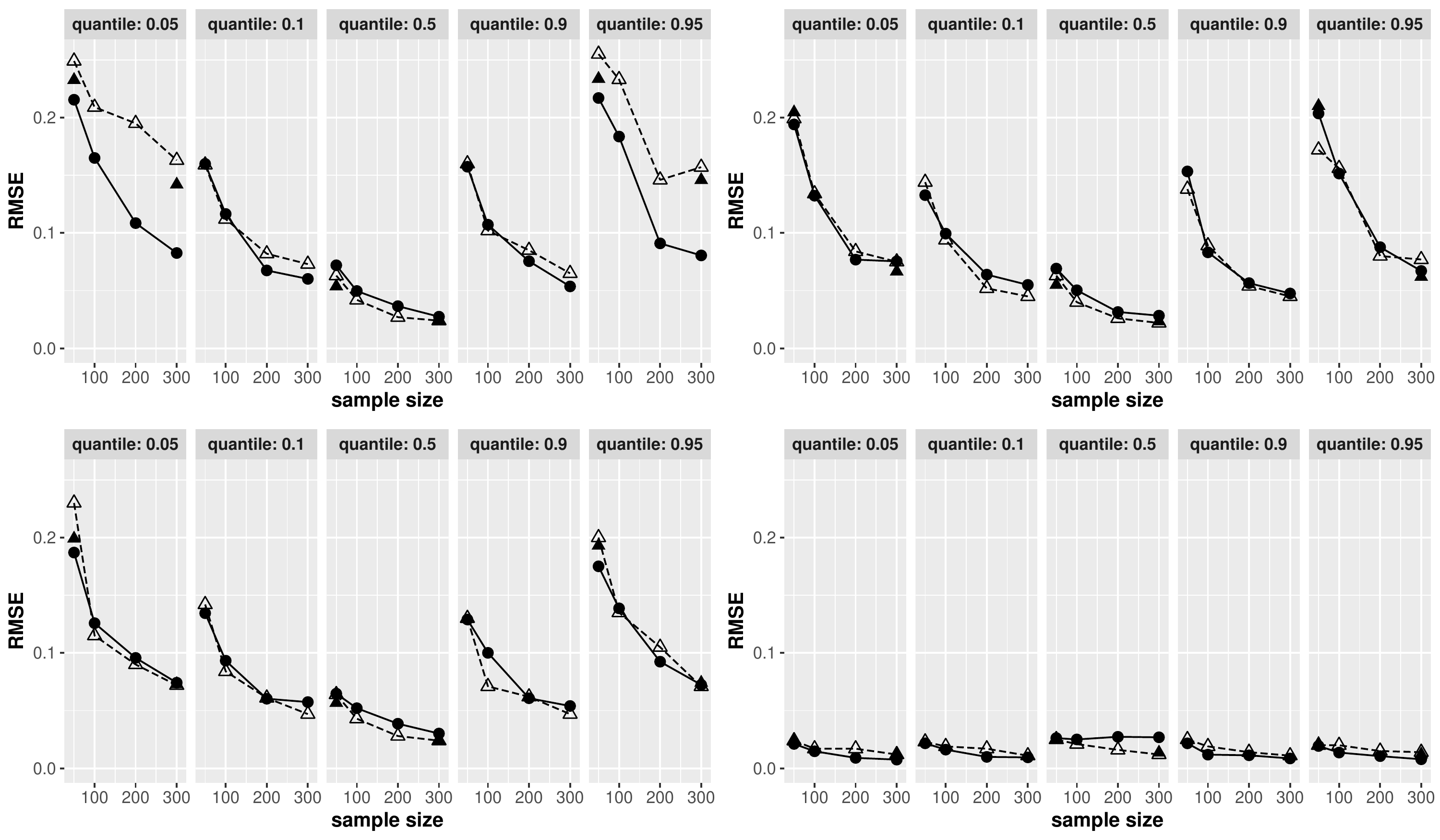}
\caption[]{Root mean squared error (RMSE) for linear quantile mixed effects estimators based on quadrature \& non-smooth optimization (filled circles) and approximated EM (filled triangles): $\beta_{\tau,0}$ (upper left panels), $\beta_{\tau,1}$ (upper right panels), $\beta_{\tau,2}$ (lower left panels), and $\sigma_{\tau}$ (lower right panels). The RMSE values reported by \citet[][Table 2]{galarza} for the approximated EM are marked with empty triangles.}
\label{fig2}
\end{figure}
\end{landscape}

Table~\ref{tab1} shows a summary of the actual performance of the two algorithms, while Figures~\ref{fig1} and \ref{fig2} show, respectively, the absolute bias and root mean squared error (RMSE) of the two estimators. For comparison, Figure~\ref{fig2} also shows the RMSE values reported for SAEM by Galarza et al in Table 2 of their paper.

The average bias and RMSE calculated for $\bm\beta_{\tau}$ and $\sigma_{\tau}$ in selected scenarios were small for both estimators (Table~\ref{tab1}), with average RMSE slightly lower for \texttt{lqmm}. Most notably, the time needed by \texttt{qrLMM} to run one model was on average 42.52 minutes (min 5.82, max 132.30 minutes) with a $21\%$ convergence failure rate. In contrast, \texttt{lqmm} took less than 1 second for one replication (min 0.09, max 9.27 seconds) with no convergence failures in any of the selected scenarios (and no convergence failures in any of the other scenarios either). The bias and RMSE for specific sample sizes and quantiles were similar in most cases except for the intercept, in which case \texttt{lqmm} seemed to have an advantage over \texttt{qrLMM} at more extreme quantiles (Figures~\ref{fig1} and \ref{fig2}). Note also that the RMSE results reported by Galarza et al in their paper are close to those we obtained in our simulation for selected scenarios (Figure~\ref{fig2}), thus it is reasonable to conclude that our simulation design is a faithful reproduction of theirs.

Table~\ref{tab2} shows that the (scaled) average log-likelihood resulting from the two fitting algorithms was comparable in all selected scenarios.

\begin{table}
\caption{Average log-likelihood at convergence (scaled by $n$) for the quadrature \& non-smooth optimization algorithm (lqmm) and the approximated EM algorithm (qrLMM) run on a 64-bit operating system machine with 32 Gb of RAM and 3.60 GHz clock-rate processor. All figures refer to the same subset of scenarios, namely $M \in\{50, 300\}$ and $\tau \in\{0.05, 0.5, 0.95\}$. Averages are calculated over $2\times 3 \times 100 = 600$ replicated datasets.}
\label{tab2}
\begin{tabular}{lrrrr}
  \hline
Algorithm & Sample size & $\tau = 0.05$ & $\tau = 0.5$ & $\tau = 0.95$ \\
  \hline
  lqmm & 50 & $-$7.78 & $-$4.14 & $-$7.77 \\
        & 300 & $-$7.83 & $-$4.28 & $-$7.83 \\
  qrLMM & 50 & $-$7.81 & $-$4.11 & $-$7.80 \\
        & 300 & $-$7.86 & $-$4.22 & $-$7.86 \\
   \hline
\end{tabular}
\end{table}

As a side note, we found Galarza el al's simulation setting rather unusual since the vector $\mathbf{z}_{ij}$ is typically a subset of $\mathbf{x}_{ij}$, for the random effects are constrained to have zero-mean (see also the discussion in the next section). Moreover, the AL distribution in LQMM provides only a quasi-likelihood for point estimation, i.e. it is not assumed to be the true distribution. Galarza el al's simulation is very limited and is more of a sanity check than a validation study. It would be more realistic to simulate errors from a variety of distributions with different shapes, along with heteroscedastic variants of these models. An extensive simulation of this kind is provided by \cite{geraci_2014b}.

\section{Framingham study}
Galarza et al also provided a comparison between the two algorithms using a subset of the cholesterol data from the Framingham study \citep{zhang}. Once again, the description of their analysis is incomplete. First of all, the authors state
\vskip 0.2cm
\begin{small}
Interestingly, for the extremes quantiles, some warnings messages on convergence were displayed while fitting Geraci's method, even after increasing the number of iterations and reducing [sic] the tolerance, as suggested in the \texttt{lqmm} manual.
\end{small}
\vskip 0.2cm
The authors do not say which estimation settings were used initially and how these were changed afterwards. Most importantly, they do not say whether the warning messages were obtained during model fitting or bootstrapping, and how many warnings were produced. These warnings may be of little concern \citep[see][for a discussion on this point]{geraci_2014a} and, as shown further below, they can be addressed with an appropriate tweaking of optimization parameters \citep{geraci_2014a}, along with a thoughtful examination of the data and model.

The authors are also silent on the model for $\bm\Sigma_{\tau}$. This is irrelevant for the \texttt{qrLMM} package since it provides only
one model (i.e., the general positive-definite matrix). However, the \texttt{lqmm} package provides four different models, including the diagonal variance-covariance structure as the default.

We then decided to replicate the analysis of Galarza et al who considered the model
\begin{equation}\label{eq:5}
Q_{y_{ij}|\mathbf{u}_{i}}(\tau) = \beta_{\tau,0} + \beta_{\tau,1}\mathrm{sex}_{i} + \beta_{\tau,2}\mathrm{age}_{i} + u_{1,i} + u_{2,i}T_{ij},
\end{equation}
where $y_{ij}$ is the $ij$th measurement of cholesterol (divided by 100) and $T_{ij} = (t_{ij} - 5)/10$, with $t_{ij}$ denoting years since the beginning of the study.

There is clearly something awkward about the specification of model \eqref{eq:5} since it does not include a fixed coefficient for $T_{ij}$. We can only surmise that Galarza et al misinterpreted equation (9) in \cite{zhang}, where the random slope is actually centered about the fixed slope (not about zero). Whatever the reason, suffice it to say that this oversight not only may introduce bias in the estimates, but it might also render estimation more difficult and prone to failure if the fixed slope is effectively different from zero. Therefore, we proceeded with the following corrected model
\begin{align}\label{eq:6}
Q_{y_{ij}|\mathbf{u}_{i}}(\tau) = & \, \beta_{\tau,0} + \beta_{\tau,1}\mathrm{sex}_{i} + \beta_{\tau,2}\mathrm{age}_{i} + \beta_{\tau,3}T_{ij}\\
\nonumber & + u_{1,i} + u_{2,i}T_{ij},
\end{align}
where $(u_{1,i}, u_{2,i})\tp \sim \mathcal{N}(\mathbf{0}, \bm\Sigma_{\tau})$ and $\bm\Sigma_{\tau}$ is a general positive-definite matrix. Model fitting and bootstrap standard error estimation were carried out for 19 vigintiles as detailed in Appendix A. The tolerance parameters set for \texttt{lqmm} estimation were less restrictive than the default values but still within reasonable bounds. We obtained only two warning messages of failed convergence during bootstrapping, but not during model estimation. Considering that $19 \times 50 = 950$ models were fitted for the bootstrap, this issue is hardly worthy of note in this case.

Finally, Galarza et al's statement
\vskip 0.2cm
\begin{small}
We observe that our SAEM method leads to mostly smaller SEs and AIC compared to the Geraci method. [...] Hence [...] the substantial gain in the AIC criterion and the SEs establish that our SAEM approach provides a much better fit to the dataset
\end{small}
\vskip 0.2cm
\noindent is a hodgepodge of claims that are not substantiated anywhere in their paper. First, the authors found in their simulation study that the asymptotic approximations provide valid standard errors \citep[][Table 1]{galarza}. This should be expected since the data were generated under the AL distribution. Theoretical results actually show that outside the AL case it is not appropriate to quantify uncertainty using this distribution as it leads to underestimation of the true variability \citep{yang}. This is why \texttt{lqmm} makes use of bootstrap. Moreover, the authors never provided a simulation study to compare SAEM's standard errors with those obtained with \texttt{lqmm}; thus, it is not possible to understand the nature of the differences found for one particular dataset (even if we grant for the sake of argument that the \texttt{lqmm}'s estimation settings and modeling syntax were appropriately specified in the Framingham data analysis). Secondly, the comparison between average log-likelihoods in our simulation (Table~\ref{tab2}) does not support the superiority of SAEM in terms of goodness of fit (GOF). In summary, the issue of standard error estimation remains to be investigated, while empirical evidence, although limited, contradicts Galarza et al's claim that SAEM gives a better GOF performance.

\section{Conclusion}
Linear quantile mixed models \citep{geraci_2014b} represent a valuable tool available to the scientific community. Computational issues are still an open problem and different approaches have been investigated by several researchers \citep[see][for an overview]{marino_2015}. Galarza et al's claim of SAEM's superior performance fails to stand up to closer examination. Our simulation shows that while SAEM produces finite-sample bias and RMSE comparable to those obtained from the quadrature-based algorithm in \texttt{lqmm}, its sluggish convergence and high proportion of convergence failures put Galarza et al's proposal at enormous disadvantage.

\appendix

\section{R code}
In this Appendix, we provide the \texttt{R} code for the analysis of the Framingham cholesterol data using the \texttt{lqmm} package.
\begin{verbatim}
library(lqmm)

data(Cholesterol, package = "qrLMM")
Cholesterol$year.c <- (Cholesterol$year - 5)/10
Cholesterol$sex <- as.factor(Cholesterol$sex)

# Set optimization parameters
ctrl <- lqmmControl(method = "df", LP_tol_ll =
1e-3, LP_max_iter = 2000, startQR = TRUE)

# Fit model for tau = 0.05, 0.1, ..., 0.95
fit <- lqmm(I(cholst/100) ~ year.c + sex + age,
random = ~ year.c, group = ID,
data = Cholesterol, tau = 1:19/20,
covariance = "pdSymm", control = ctrl)

# Bootstrap (50 replicates)
fit.s <- summary(fit, R = 50, seed = 178)
\end{verbatim}

\section*{Acknowledgements}
This research was partially supported by the National Institutes of Health -- National Institute of Child Health and Human Development (Grant Number: 1R03HD084807-01A1).


\end{document}